# Large magnetoresistance in the magnetically ordered state as well as in the paramagnetic state near 300 K in an intermetallic compound, Gd$_7$Rh$_3$

*Kausik Sengupta, S. Rayaprol and E.V. Sampathkumaran*
*Tata Institute of Fundamental Research, Homi Bhabha Road, Colaba, Mumbai-400005*

**Abstract**
*We report the response of electrical resistivity ρ to the application of magnetic fields (H) up to 140 kOe in the temperature interval 1.8-300 K for the compound, Gd$_7$Rh$_3$, ordering antiferromagnetically below 150 K. We find that there is an unusually large decrease of ρ for moderate values of H in the close vicinity of room temperature uncharacteristic of paramagnets, with the magnitude of the magnetoresistance increasing with decreasing temperature as though the spin-order contribution to ρ is temperature dependent. In addition, this compound exhibits giant magnetoresistance behaviour at rather high temperatures (above 77 K) in the magnetically ordered state due to a metamagnetic transition.*

**INTRODUCTION**

In the field of giant magnetoresistance (GMR), the GMR behaviour is usually observed around (or below) respective magnetic ordering temperatures ($T_o$) of materials and is related to a variety of physical phenomena (like double-exchange, metal-insulator transition, charge ordering etc., see, for instance, [1-3]) setting in around this temperature (T). In the paramagnetic state of a magnetic material without such phenomena, far above $T_o$, the magnitude of the magnetoresistance (MR) is usually negligible, particularly as the room temperature range is approached. Here, we report significant temperature dependent MR effect in the close vicinity of 300 K in the paramagnetic state of a rare-earth (R) intermetallic compound. This is observed for the polycrystalline form of a member (R= Gd) [4] ordering antiferromagnetically below 140 K, in the family [5] $R_7Rh_3$, crystallizing in $Th_7Fe_3$-type hexagonal structure (space group $P6_3mc$). In addition, we observe GMR behaviour in the magnetically ordered state as well, particularly beyond a certain value of magnetic field (H), which is attributed to a metamagnetic transition.

**EXPERIMENTAL DETAILS**

The compound, Gd$_7$Rh$_3$, in the polycrystalline form was prepared by arc melting stoichiometric amounts of high purity (> 99:9%) Gd and Rh in an arc furnace in an inert atmosphere. The ingot was subsequently annealed in an evacuated sealed quartz tube at 300 $^0$C for 50 hours. The sample thus prepared was found to be single phase by x-ray diffraction. The sample was further characterized by temperature dependent magnetization (M) measurements in a commercial (Oxford Instruments) vibrating sample magnetometer, particularly by the observation of $T_o$ at 140 K, in agreement with the literature [4]. Isothermal M measurements were performed at selected temperatures up to 120 kOe with the same magnetometer. The ρ behaviour in the temperature range 1.8-300 K up to H= 140 kOe was measured in a commercial (Quantum Design) Physical Property Measurements System by a conventional four-probe method employing a conducting silver paint for making electrical contacts of the leads with the sample.

**RESULTS**

In Fig. 1a, we show the ρ(T) behaviour in the presence of various external magnetic fields. Though this is an intermetallic compound, in zero field, the sign of the temperature coefficient (dρ/dT) of ρ is found to be negative above 140 K instead of a positive sign expected for metals, which by itself makes this compound interesting. Considering that the ρ values are rather large, one is tempted to attribute this negative sign to a weak-localization effect arising from disorder and/or complicated crystal structure (with three inequivalent sites for R). Though the presence of such an effect to some extent can not be ruled out completely, we believe that it can not be the sole explanation for negative dρ/dT as many light R members of this rare-earth series exhibit [6] a positive sign of dρ/dT. Even some heavy R members like Dy and Ho show [4, 6] a positive sign of dρ/dT at least in a narrow temperature range above $T_o$, followed by a negative sign at still higher temperatures resulting in a maximum in ρ(T) in the paramagnetic state. Therefore, the existence of band gap was proposed [4, 6] for the temperature range in which the sign of dρ/dT is negative. We will address this issue further later in this article.

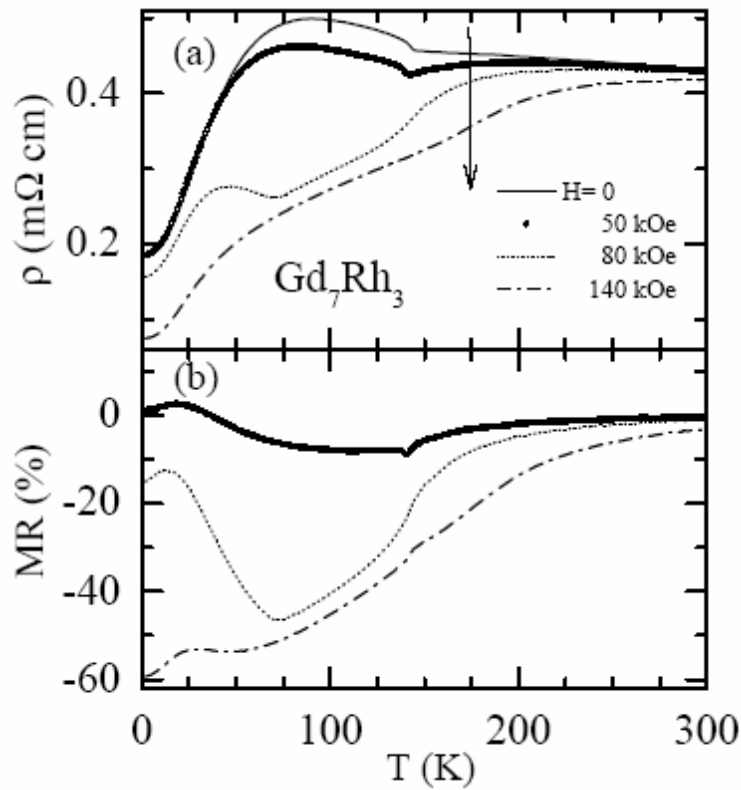

**Fig 1** (a) Electrical resistivity (ρ) in the presence of externally applied magnetic fields (H) and (b) the magnetoresistance (MR), as a function of temperature (T) for $Gd_7Rh_3$.

Below 140 K, there is an upturn which is attributed to the formation of magnetic Brillouin-zone boundary gaps [7] due to the antiferromagnetic nature of the magnetic transition, followed by a broad maximum around 100 K. Apparently, these gaps tend to diminish as T is lowered towards 1.8 K, as indicated by the positive sign of dρ/dT. These features in ρ(T) are in good agreement with those reported in the literature [4].

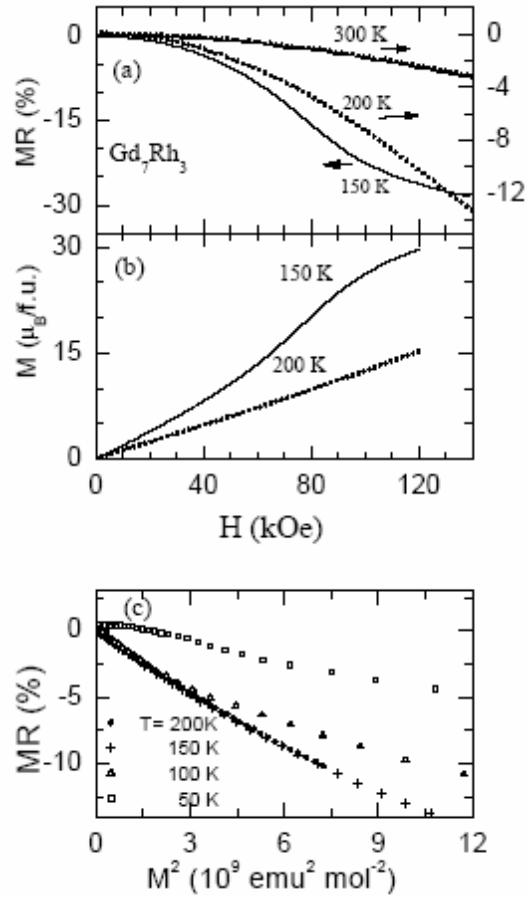

**Fig. 2** (a) Magnetoresistance (MR) and (b) isothermal magnetization (M) as a function of H, at 150, 200 and 300 K for $Gd_7Rh_3$. The scaling of MR with $M^2$ in the paramagnetic state is demonstrated in (c) by choosing the data at 150 and 200 K; its failure in the magnetically ordered state is also demonstrated in this figure using the M data shown in figure 3.

Let us now look at the influence of externally applied fields on these features. It is clear from Fig. 1a that the temperature range over which the negative sign of dρ/dT is observed above 140 K diminishes with a gradual increase of H resulting in a positive dρ/dT in a range above 140 K as well. That is, there is a maximum in ρ(T) appearing in the paramagnetic state, say, around 200 K, for H = 50 kOe. It is to be remarked that there is a noticeable decrease in the value of ρ in the presence of H. In order to get a better idea of the magnitude of the change in ρ with H, we have plotted in Fig. 1b the values of MR, defined as [ρ(H)- ρ(0)]/ρ(0), obtained from the data shown in Fig. 1a. In addition, we show in Fig. 2a the values of MR obtained by measuring ρ as a function of H at fixed temperatures. It is to be noted (see Fig. 2a) that, even at 300 K, the magnitude of MR observed at high fields is significant, for instance, about -2% and -3% for H= 100 and 140 kOe respectively. The finding of central importance is that, as the T is lowered, say to 200 K, the magnitude of MR grows further (see also Fig. 1b), say to -6:5% and -12:5% respectively; at 150 K, the corresponding values are still larger (about -20% and -30% respectively); it is clear from the figures 1b and 2a that the magnitudes are large even for relatively moderate values of H, say 50 kOe. Such a large decrease of ρ with H in this high temperature range is unusual for intermetallic compounds in the paramagnetic state. In metals [8], the negative contribution due to the suppression of paramagnetic fluctuations by H in this temperature range is negligible or comparable to the positive contribution to MR at high temperatures from the influence of H (quadratic field dependence) on the conduction electron motion, resulting in negligible net MR. We measured several other paramagnetic Gd compounds (containing similar atomic ratio of Gd as well as those with different residual resistivity ratios) and this is found to be the case [9].

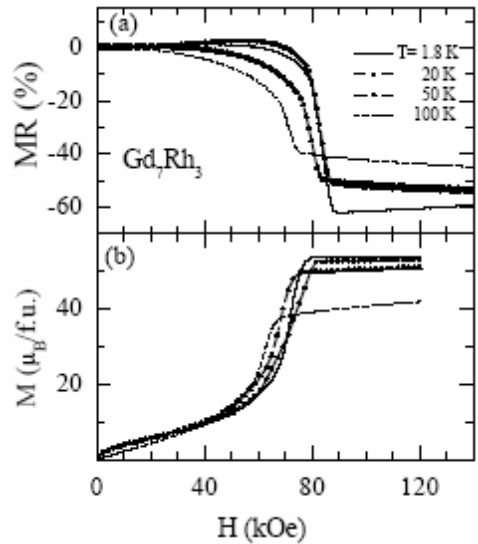

**Fig. 3** (a) Magnetoresistance (MR) and (b) isothermal magnetization (M) as a function of H, at selected temperatures in the magnetically ordered state of $Gd_7Rh_3$.

**DISCUSSION**

In what follows, we will discuss possible explanations for the GMR behaviour in the paramagnetic state.

On the basis of the ρ(T) behaviour (in zero field) of this family of compounds, it was suggested [4, 6] that the light R members exhibit metallicity in the entire temperature range, whereas semimetallic band-structure develops with increasing T as the ionic radius of R diminishes. One is therefore tempted to propose that the large MR in the present material may arise from the influence of H on the semimetallic energy gap (with a corresponding increase in the carrier density). But this factor, if contributes, can not be the sole cause of the observed MR anomaly, as we recently found [10] large values (but of relatively less magnitude) of MR for the Dy analogue even in a temperature range (in the paramagnetic state, just above $T_o$) in which zero-field dρ/dT is positive; in addition the magnitude of MR at 300 K (at which the band-gap was proposed to be present [4, 6]) even at H = 140 kOe is found to be less than 0.1% for this Dy compound. Therefore, one has to search for alternate explanations. Since the feature emphasized is pertinent to the paramagnetic state, we have to propose that there are unusually strong paramagnetic scattering effects [11, 12] which are suppressed by the presence of H as indicated by the negative sign of MR. In support of this magnetism-related explanation, MR at 150 and

200K scales [13] (see Fig. 2c) with the square of isothermal M (shown in Fig. 2b) (though it breaks down below 140 K presumably due to the formation of magnetic gaps). This finding implies that the spin-disorder contribution is apparently large. Therefore, we speculate that the large absolute values of $\rho$ could be intrinsic, and hence need not be associated with any lattice disorder effect. Judged by increasing magnitude of MR with decreasing T, this contribution is apparently temperature dependent in contrast to the traditional belief that this should be a constant in the paramagnetic state for Gd (considering that the crystal-field effects have to be ignored in the entire temperature range of investigation). Possible root-cause of this enhanced spin-disorder effect is addressed further below.

It may be recalled that the Kondo effect in the paramagnetic state (at low temperatures) also can give rise [14] to large magnitudes of MR and temperature dependent spin-disorder contribution. However, it is a well-known fact that such a phenomenon is not possible for Gd, as the moment-carrying 4f orbital is deeply localized. There is no evidence till to date for the moment on Rh for non-magnetic (or light) rare-earth members in this family. If there is a small 4f-induced moment on Rh in the Gd compound, the spatial extension of Rh-4d orbitals should favour itinerant magnetism, thus acting against any possible Kondo effect for which localized nature of the magnetic moment is a prerequisite. It is not clear whether anomalously large paramagnetic fluctuations within such an itinerant band of Rh 4d is responsible for large MR even around 300 K. In this connection, it is worth noting that, though the effective magnetic moment inferred from our magnetic susceptibility data (not reported here) in the paramagnetic state essentially corresponds to that of free Gd ions alone, we observe an excess value of saturation moment (about 1 $\mu_B$/Rh) at 1.8 K (see Fig. 3b), which signals the existence of an induced moment in the 4d band of Rh at very low temperatures. Hence it will be interesting to focus future studies on the role of Rh 4d band on the transport anomalies. However, since few other Gd alloys exhibiting similar transport anomalies [11, 12] do not contain such 4d-metals, we are more in favor of attributing such MR anomalies to a hither-to-unexplained magnetic precursor effect [11, 12] as a common mechanism in all these compounds. Incidentally, the present observation is unique considering that such previous reports [11-13] are pertinent to temperatures below 100 K only. We would like to add that it is of interest to perform MR studies on single crystals to explore whether there is any unusual role of grain-boundary effect in the paramagnetic state.

We now present MR behaviour in the magnetically ordered state. At very low temperatures (say, 1.8-20 K), the sign of MR is positive (Fig. 1b) for moderate applications of H (till about 65 kOe), typical of antiferromagnetic metals without magnetic gap effects; however, as T is increased towards $T_o$, MR tends to change the sign (see also the curves for 50 and 100 K in Fig. 3a) for such moderate fields; this could be attributed to the influence of H on the magnetic gapped Fermi surface for the intermediate temperature range below To. The most notable finding is that there is a sharp increase (Fig. 3a) in the magnitude of MR with increasing H (taken at fixed temperatures) in the vicinity of 70-90 kOe, presumably due to a sudden change in the type of magnetic ordering, that is, from antiferromagnetism to ferromagnetism. The isothermal M data shown in Fig. 3b endorses this conclusion and the value of saturation moment, say at 1.8 K and H= 80-120 kOe, is even marginally higher than that expected for ferromagnetically coupled Gd ions (7 $\mu_B$/Gd). The field at which this crossover takes place decreases marginally with increasing T (by about 10 kOe as T is varied from 5 K to 80 K). Interestingly, the magnitude of MR becomes huge following the spin-reorientation transition. Thus, even at temperatures as high as 100 K, we observe a large value for MR, for instance about -40% around 80 kOe. The magnitude increases to as much as -60% at 2 K for H= 80 kOe. Thus, this compound belongs to a small class of intermetallic materials in which GMR behaviour has been seen at rather high temperatures due to metamagnetic-like transitions (see, for instance, Refs 15-18).

**CONCLUDING REMARKS**

To summarise, the compound $Gd_7Rh_3$ exhibits large magnetoresistance not only in the magnetically ordered state (below 140 K), but also over a wide temperature range in the paramagnetic state (even in the vicinity of room temperature) far above Néel temperature. While this work demonstrates an unconventional strategy with which one can engineer GMR effects near room temperature, particularly among intermetallics, it is a theoretically challenging task to understand how spin-disorder contribution becomes apparently large and also temperature dependent in the paramagnetic state of some antiferromagnetic intermetallic compounds. However, in the case of ferromagnets, there is a recent theoretical attempt [19] to understand such GMR anomalies in terms of critical point effects. Finally, it is also worth noting that the scaling between $M^2$ and MR discussed in Ref. [20] for many other GMR systems in the magnetically ordered state interestingly is found to break down in the present compound below Néel temperature (see Fig. 2c). Thus, this work brings out that such a relationship is not applicable for antiferromagnets, particularly when there is a magnetic gap formation.

**Acknowledgements**

We thank Kartik K. Iyer for his help during measurements.